# 3D PIC simulation and theoretical modeling of RF-laser pulse in magnetized plasma for the generation of multidimensional relativistic Wakefields

A. A. Molavi Choobini[1] and M. Shahmansouri[1*]

[1]Department of Atomic and Molecular Physics, Faculty of Physics, Alzahra University, Tehran, Iran.

**Abstract:**

   The present study, investigates the modulation of plasma wakefields in dense magnetized plasma driven by relativistic electron beams under transverse RF excitation. A self-consistent theoretical framework, comprising the RF vector potential, Maxwell's equations, and relativistic electron motion, is extended through full 3D electromagnetic particle-in-cell simulations. The results reveal systematic amplification and reshaping of wakefields under the combined action of external magnetic fields and RF drivers. Variations in the cyclotron-to-plasma frequency ratio dictate the radial positioning and gyromotion of plasma electrons, sharpening transverse confinement and stabilizing blowout structures. The RF amplitude introduces progressive modulation of radial excursions and transverse forces, enhancing wakefield symmetry and depth. Current density distributions confirm the nonlinear scaling with RF strength, evolving from weak perturbations into sharply structured ion channels. Scalar potentials and longitudinal fields exhibit pronounced sensitivity to pulse shape, polarization angle, frequency ratio, and driver density, each parameter producing distinct oscillatory features and confinement regimes. Plasma density sets the field strength and radial localization, while the modulation parameter $\kappa$ governs the emergence of fine-scale oscillatory bands, producing smooth-to-multiband transitions in longitudinal electric fields. Across all conditions, simulations confirm the reinforcement of ponderomotive force, resulting in controlled narrowing of electron sheaths, sharper scalar potential gradients, and extended acceleration zones.

**PACS:** 52.38.Kd, 41.60.Cr, 52.65.Rr.

**Keywords:** 3D PIC simulation, Laser-plasma interactions, plasma accelerator, RF-laser pulse, magnetized-plasma, wakefields.

## I. Introduction

   In plasma-based wakefields, an intense particle or laser beam drives a plasma wave with a phase velocity near the speed of light. When a laser beam is employed, the process is termed laser wakefield acceleration (LWFA) [1-3]. Conversely, when a particle beam is used, it is referred to as plasma wakefield acceleration (PWFA) [4, 5]. In LWFA, a short-pulse, high-intensity laser is directed into a plasma, wherein the electric field of the laser induces a wake of plasma oscillations behind it [6, 7]. Electrons within the plasma can "surf" on these waves, thereby gaining energy. In PWFA, a high-energy particle beam, typically consisting of electrons or positrons, drives the wakefield in the plasma [8, 9]. This particle beam generates a wake of plasma oscillations, which subsequent particles can utilize for acceleration. While LWFA and PWFA predominantly operate in the GHz frequency regime, recent research has been exploring the potential for these wakefield accelerators to function in the THz frequency regime [10-12]. This advancement entails generating and utilizing plasma waves at THz frequencies, which can provide higher acceleration gradients and enable more compact accelerator structures. Operating in the THz frequency regime for both LWFA and PWFA could significantly enhance the efficiency and capabilities of these accelerators.

Recent advances in laser–plasma interactions have deepened understanding of nonlinear dynamics and wakefield excitation in relativistic regimes. Ping-Tong Qian et al. investigated the nonlinear interaction of relativistic pulses with hot plasma, showing that density and thermal effects drive soliton stability and modulation instability [13]. Yuan Shi and colleagues extended strong-field plasma theory to include three-wave interactions and transitions from wakefield acceleration to electron–positron pair production [14]. Complementary to these studies, Martin King and colleagues reviewed relativistic transparency in near-solid targets with multi-petawatt lasers, highlighting enhanced absorption, novel acceleration pathways, and structured radiation [15]. Modeling techniques are crucial for scaling to practical accelerators. D. Terzani and team validated the time-averaged ponderomotive approximation for long-distance LWFA, confirming accuracy with reduced computational cost [16], while F. Massimo and co-workers combined it with Lorentz-boosted frames to simulate meter-scale, multi-GeV accelerators and assess feasibility of multi-TeV colliders [17]. In the nonlinear bubble regime, Golovanov and Kostyukov developed a planar 2D model for non-uniform plasma, supported by PIC simulations [18]. For diagnostics, Yadav et al. modeled betatron radiation with Liénard–Wiechert potentials and QED benchmarks, accurately predicting angular–energy spectra for PWFA experiments [19]. Laser pulse engineering has proven effective for enhancing LWFA. Hafz and colleagues demonstrated dual-color pulses increasing electron energies to 700–800 MeV [20], while Tomkus and team showed that Bessel–Gauss beams extend acceleration length and boost energies [21]. Khudiakov and Pukhov proposed an optimized injection scheme via femtosecond laser–solid interactions, achieving GeV-scale bunches with <1% energy spread [22]. Sedaghat and co-workers further revealed that positively skewed pulses nearly double energy gain, albeit with stronger transverse fields [23]. Beam-driven and radiation-focused approaches provide complementary routes. Sandberg and Thomas proposed a PWFA scheme with tapered density profiles to generate ultrashort XUV pulses [24]. Wan et al. directly visualized wakefield dynamics with femtosecond electron probing [25]. Lamač and colleagues reported long-lived magnetic fields up to $10^{10}$ G in underdense plasma, with decay governed by filament instabilities [26].

Most existing theories on wakefield excitation have been limited to linear fluid theory or one-dimensional nonlinear fluid theory. Nevertheless, recent PWFA and LWFA experiments have shown wake excitation under blowout conditions, in which neither fluid dynamics nor one-dimensional theory can properly describe the physics. In this study, a predictive theoretical model for wake excitation by RF-laser pulses in the blowout regime is introduced, synergistically combining relativistic electron beam-driven plasma wakes with externally imposed transverse RF fields and uniform magnetic fields to achieve unprecedented control over wakefield amplitude, transverse focusing, and phase-locking in dense plasmas. This validated by electromagnetic 3D particle-in-cell simulations that capture multidimensional oscillations beyond quasistatic approximations. Through transverse RF modulation, dynamic manipulation of electron radial dynamics and ponderomotive forces is realized, enabling precise tuning of blowout regime characteristics, including sheath narrowing and ion channel stabilization. Resonant interactions between RF fields and plasma oscillations, governed by cyclotron-to-plasma frequency ratios, amplify gyromotion effects, confining trajectories within Larmor radii and producing tightly structured wakefields with enhanced longitudinal electric fields. This resonant enhancement establishes levels of wakefield uniformity and amplitude scaling unattainable in conventional unmodulated plasma wakefield accelerators. Elliptical RF polarization, particularly with phase offsets, facilitates controlled transitions between symmetric and asymmetric wake structures, modulating scalar potential peaks and longitudinal field asymmetries to optimize phase-matching and energy transfer efficiency. Variations in driver pulse shapes—such as Gaussian for sharp, symmetric ion channels, chirped-Gaussian for asymmetric phase velocities, ring-shaped for broadened charge separations, and cosh-Gaussian for extended oscillatory tails—generate tailored wakefield excitation profiles, revealing dependencies absent in standard blowout models. Systematic adjustments in RF amplitudes, electron beam spot sizes, and beam-to-plasma density ratios intensify nonlinear plasma responses, leading to quadratically scaled ponderomotive-driven expulsions that deepen scalar potential wells, extend longitudinal field oscillations, and broaden acceleration zones while maintaining peak gradients. The introduction of modulation parameters such as κ drives the emergence of highly oscillatory multi-band structures in longitudinal fields, evolving from smooth single-peaked distributions in unmodulated regimes to densely packed alternating bands that signify strong RF–plasma synchronization. The presented framework, supported by analytical derivations of electron dynamics, charge continuity, and relativistic motion under RF and magnetic influences, establishes a predictive regime for high-gradient, stable wakefield acceleration. The approach advances the precision, coherence, and scalability of plasma-based colliders by leveraging synchronized gyromotion, RF-induced phase matching, and enhanced polarization control in ultra-relativistic dense magnetized plasmas. The paper is organized as follows: Section II presents the theoretical framework for RF-modulated plasma dynamics. Section III details the numerical simulations and validation. Section IV discusses parameter dependencies and wakefield enhancements. Conclusions are drawn in Section V.

## II. Theoretical Model

Consider a highly relativistic electron ($\gamma \gg 1$) confined within a magnetized plasma, characterized by the longitudinal coordinate $\xi = k_p(z - \beta_{ph}ct)$, a dimensionless co-moving variable, where $\beta_{ph}$ represents the phase velocity normalized to the speed of light $c$. The radio-frequency (RF) vector potential described as a transverse spatiotemporal wave, is expressed as:

$$\vec{A}_{RF}(z,t) = A_{0x}^{RF} \cos(k_{RF}z - \omega_{RF}t)\hat{e}_x + A_{0y}^{RF} \cos(k_{RF}z - \omega_{RF}t + \varphi)\hat{e}_y \quad (1)$$

here $A_{0x}^{RF}$ and $A_{0y}^{RF}$ denote the amplitudes of the RF vector potential along the $x-$ and $y-$axes, $k_{RF}$ is the RF wavenumber, $\omega_{RF}$ is the angular frequency, and $\varphi$ represents the phase offset between the field components, respectively. The phase $\varphi$ facilitates elliptical polarization, which governs the transverse dynamics of the electron beam and the polarization properties of the emitted radiation. Furthermore, a uniform external magnetic field $\vec{B}_{ext} = B_0 \hat{e}_z$ is imposed along the direction of transverse RF electric field that drives the wake excitation. To analyze multidimensional plasma oscillations excited by an intense electron beam augmented by a transverse RF electric field for phase-locked driving, employing Maxwell's equations in the Lorentz gauge along with the equation of motion is considered as:

$$\left(\frac{1}{c^2}\frac{\partial^2}{\partial t^2} - \nabla^2\right)\vec{A} = -\frac{4\pi}{c}\vec{J} \tag{2a}$$

$$\frac{1}{c}\frac{\partial \Phi}{\partial t} + \vec{\nabla}\cdot\vec{A} = 0 \tag{2b}$$

$$\frac{d\vec{P}_e}{dt} + \vec{V}_e\cdot\vec{\nabla}\vec{P}_e = -e\left(\vec{E} + \frac{\vec{V}_e}{c}\times\vec{B}\right) + \vec{F}_{pond} \tag{2c}$$

where $\vec{E} = \vec{E}_p + \vec{E}_{RF}$ includes the plasma fields ($\vec{E}_p$) and the RF driver ($\vec{E}_{RF} = -\partial\vec{A}_{RF}/\partial t$). While $\vec{B} = \vec{B}_p + \vec{B}_{ext}$ incorporates induced fields and the external field, introducing gyration that stabilizes blowout in dense plasmas, and $\vec{J} = \sum_e -en_e\vec{V}_e$. The total ponderomotive force $\vec{F}_{pond}$ is expressed as:

$$\vec{F}_{pond} = \vec{F}_{pond,Laser} + \vec{F}_{pond,RF} = \frac{e}{2}\vec{\nabla}[\bar{\gamma}_e m_e c^2(|A_L|^2 + |A_{RF}|^2)] \tag{3}$$

where $\bar{\gamma}_e = \sqrt{1 + \frac{P_e^2}{m_e^2 c^2} + \frac{e^2 B_0^2 r_L^2}{m_e^2 c^4} + \frac{e^2|A_L|^2}{2m_e^2 c^4} + \frac{e^2|A_{RF}|^2}{2m_e^2 c^4}}$ via Larmor radius $r_L = \gamma m_e v_\perp c/eB_0$. The ponderomotive term in Eq. (3) represents the cycle-averaged force arising from the gradient of the oscillatory electron kinetic energy. Through averaging, the rapidly oscillating components reduce to the gradient of an effective potential proportional to the squared field amplitudes, yielding separate contributions from the laser and RF drivers. This formulation relies on two considerations: (i) for widely separated and incoherent frequencies, cross terms vanish and the responses add linearly; (ii) for phase-locked or nearly resonant fields, residual beat terms can survive, producing low-frequency modulations that alter the plasma response. In the strongly relativistic regime ($\gamma \gg 1$, $A_L \gg A_{RF}$), the laser dominates while the RF field acts as a perturbation, whereas near plasma resonance, even a modest RF amplitude can phase-lock the wake and improve energy transfer. An external magnetic field further modifies the effective ponderomotive force through $\bar{\gamma}_e$, where gyromotion redistributes oscillatory energy and introduces directional dependence. Within the quasistatic, or frozen-field, approximation, the driver evolves on a timescale significantly longer than that of the plasma response. As a result, its temporal shape remains nearly unchanged while passing through the plasma electrons, and the driver can therefore be treated as stationary when evaluating the plasma response. To utilize the quasistatic approximation, transformation of the variables ($\partial/\partial t \ll \partial/\partial \xi$) is implied and Maxwell's equations then become as follows:

$$E_z = \frac{\partial \Phi}{\partial \xi}, \qquad \vec{E}_\perp = -\vec{\nabla}_\perp \Psi - \frac{\partial \vec{A}_\perp}{\partial \xi} + \vec{E}_{RF} \tag{4a}$$

$$B_z = \frac{\partial}{\partial \xi}(\vec{\nabla}_\perp \cdot \vec{A}_\perp), \qquad \vec{B}_\perp = \vec{\nabla}_\perp \Psi \times \hat{e}_z + \vec{\nabla}_\perp \times \vec{A}_\perp + \vec{B}_{ext} \tag{4b}$$

$$\nabla_\perp^2 \vec{A} = -\vec{J}, \qquad \nabla_\perp^2 \Phi = -\rho, \qquad \vec{\nabla}_\perp \cdot \vec{A}_\perp = -\frac{\partial \Psi}{\partial \xi} \qquad (4c)$$

here $\Psi = \Phi - A_z$ ($\Phi$ is the scalar potential and $A_z$ is the axial vector potential component), $\vec{A}_\perp = A_x \hat{e}_x + A_y \hat{e}_y$ and $\vec{\nabla}_\perp = \frac{\partial}{\partial x}\hat{e}_x + \frac{\partial}{\partial y}\hat{e}_y$. Therefore, the plasma electrons evolve as:

$$\frac{d\vec{P}_\perp}{d\xi} = \frac{1}{1-v_z}\left[-\vec{\nabla}_\perp \Phi + (\vec{V}_\perp \times \vec{B})_z - \frac{e^2}{2\bar{\gamma}_e m_e^2 c^4}\vec{\nabla}_\perp(|A_L|^2 + |A_{RF}|^2)\right] - \frac{\omega_c}{\omega_p}v_\theta \qquad (5)$$

Hence, the wakefield $E_z$ can be completely expressed in terms of the potential $\Phi$, which derived from the Poisson equation ($\vec{\nabla}_\perp \Phi = J_z - \rho$). The source term is determined by the plasma charge density ($\rho$) and longitudinal current ($J_z$), which themselves are coupled to all fields via the equation of motion. The inclusion of $\vec{A}_{RF}$ and $\vec{B}_{ext}$ enhances the model's applicability to dense magnetized plasmas, where the blowout regime benefits from improved transverse focusing and higher wake amplitudes due to RF phase-matching and magnetic confinement. The $v_\theta$ represents the azimuthal (angular) velocity component of the plasma electrons due to their gyromotion in the presence of the external magnetic. In the blowout regime, the plasma wake excitation in a dense magnetized plasma is driven by either an intense electron beam, where the plasma electrons are predominantly expelled transversely, forming a narrow ion channel. The key assumption is that all electrons within the blowout radius $r_b(\xi)$ are completely expelled, forming a thin sheath just beyond $r_b$, surrounded by a weakly responding (linear) region. This holds when the beam or laser spot size is much smaller than $r_b$, and in dense magnetized plasmas, the magnetic field narrows the sheath thickness to $\sim r_L$, enhancing uniformity. The expelled electrons return due to ion space charge, creating the wake. For this, assume a dual Gaussian electron driver as a density profile as follows:

$$n_b(r,\xi) = \frac{N}{\sqrt{(2\pi)^3}\Theta_r^2 \Theta_z^2} exp\left(-\frac{r^2}{2\Theta_r^2}\right) exp\left(-\frac{\xi^2}{2\Theta_z^2}\right) \qquad (6)$$

By assuming azimuthal symmetry, the components of vector potential and scaler potential inside the ion channel, i.e., for $r \ll \Theta_r$ and $r \leq r_b$ can be obtained as follows in cylindrical coordinate:

$$\Phi(r,\xi) = \Phi_0(\xi) + \frac{E_{0x}^{RF}Sin\phi_{RF}}{m_e\omega_p}rCos\theta + \frac{E_{0y}^{RF}Sin(\phi_{RF}+\varphi)}{m_e\omega_p}rSin\theta - \frac{r^2}{4} + \Lambda(\xi)Lnr \qquad (7a)$$

$$A_z(r,\xi) = A_{0z}(\xi) + \Lambda(\xi)Lnr \qquad (7b)$$

$$A_r(r,\xi) = \Lambda(\xi)r \qquad (7c)$$

where $\Lambda(\xi) = \int_0^r r' n_b dr'$, $\phi_{RF} = k_{RF}z - \omega_{RF}t$, $A_{0z}(\xi) = A_{0z}(r=0,\xi)$, and $\Phi_0(\xi) = \Phi_0(r=0,\xi)$. The RF field adds a transverse oscillatory term to the fields, that radially is $E_{RF,r} = \hat{e}_r \cdot \vec{E}_{RF}$. Since, the beam is highly relativistic, the force on a plasma electron at $r_b$, can be written as:

$$F_\perp = -(E_r - v_z B_\theta) + E_{RF,r} - \frac{\omega_c}{\omega_p}v_\theta$$

$$= -\frac{r}{2} + \frac{(1-v_z)\Lambda(\xi)}{r} + (1-v_z)\Lambda(\xi)\frac{dr}{d\xi} + E_{0x}^{RF}Sin\phi_{RF}Cos\theta + E_{0x}^{RF}Sin(\phi_{RF}+\varphi)Sin\theta - \frac{\omega_c}{\omega_p}v_\theta \qquad (8)$$

Equation (8) expresses the transverse force acting on electrons at the blowout boundary ($r_b(\xi)$) as a combination of the restoring electrostatic term from the evacuated ions, the longitudinal current contribution $\Lambda(\xi)$, and the RF-induced ponderomotive component. To close this force relation, one must determine the scalar potential $\Phi$ that mediates the restoring field. This is accomplished by invoking the quasi-static continuity condition ($\frac{\partial}{\partial \xi}(\rho - J_z + J_{z,RF}) + \vec{\nabla}_\perp \cdot \vec{J}_\perp = 0$) for charge and current: longitudinal variations of charge imbalance are coupled to transverse currents $J_\perp$, which in turn act as the source term for Poisson's equation. Under cylindrical symmetry and the thin-sheath approximation near $r_b(\xi)$, the Poisson equation can be solved by radial integration, linking $\Phi(r,\xi)$ explicitly to $\Lambda(\xi)$. This step provides the self-consistent connection between the macroscopic force balance in Eq. (8) and the microscopic charge distribution that sustains the wakefield. Due to this, the Poisson equation is:

$$-\frac{1}{r}\frac{\partial}{\partial r}\left(r\frac{\partial \Phi}{\partial r}\right) = \rho - J_z + J_{z,RF} \tag{9}$$

with integration, use initial condition, and boundary matching, the scaler potential is obtained as:

$$\Phi(r,\xi) = \frac{1}{4}\left[\left(A_{0x}^{RF} Cos\left(\frac{k_{RF}}{k_p}\xi\right)\right)^2 + \left(A_{0y}^{RF} Cos\left(\frac{k_{RF}}{k_p}\xi + \varphi\right)\right)^2\right] + \frac{r^2(\xi)}{4}\left[1 + \kappa(\xi)e^{-r^2/r_b^2}\right] \tag{10}$$

here $\kappa(\xi)$ is a dimensionless parameter that modulates the strength of the potential. The scalar potential given in Eq. (10) combines a baseline parabolic term ($r^2/4$ and $\Lambda(\xi)$-dependent contributions) with oscillatory corrections induced by the RF driver. To translate this potential into an explicit electron equation of motion, one evaluates its radial and longitudinal gradients and substitutes them into the relativistic transverse momentum equation. In this process, the cycle-averaged RF contribution enters, while the cyclotron correction appears through the dimensionless factor ($\omega_c/\omega_p$), arising from the Lorentz term associated with gyromotion. Therefore, the differential equation of motion for plasma electrons given by:

$$\frac{d}{d\xi}\left[(1 + \Phi(r,\xi))\frac{dr}{d\xi}\right]$$
$$= r\left[-\frac{1}{4}\left(1 + \frac{1}{1+\Phi(r,\xi)} + \left(\frac{dr}{d\xi}\right)^2\right) + \frac{d\Phi(r,\xi)}{d\xi} + \kappa(\xi)e^{-r^2/r_b^2} + \langle E_{RF,r}\rangle - r\left(\frac{\omega_c}{\omega_p}\right)^2\right] \tag{11}$$

This equation captures a balance between electrostatic restoring forces, the RF-induced ponderomotive drive, and the magnetic confinement term. If assume the dimensionless parameter ($\kappa$) and the width of the linear region with respect to the co-moving longitudinal coordinate ($\xi$) are very small and nearly negligible, the differential equation reduces to:

$$\Gamma_1(r)\frac{d^2r}{d\xi^2} + \Gamma_2(r)r\left(\frac{dr}{d\xi}\right)^2 + \Gamma_3(r)r = \frac{\Lambda(\xi)}{r} + \langle E_{RF,r}\rangle r - \left(\frac{\omega_c}{\omega_p}\right)^2 r^2 \tag{12}$$

where

$$\Gamma_1(r) = 1 + \left[\frac{1}{4} + \frac{\kappa}{2} + \frac{1}{8}r\frac{\partial \kappa}{\partial r}\right]r^2 \tag{13a}$$

$$\Gamma_2(r) = \frac{1}{2} + \frac{3}{4}\kappa + \frac{3}{4r}\frac{\partial \kappa}{\partial r} + \frac{1}{8r^2}\frac{\partial^2 \kappa}{\partial r^2} \tag{13b}$$

$$\Gamma_3(r) = \frac{1}{4}\left[1 + \frac{1+\frac{1}{2}(|A_L|^2+|A_{RF}|^2)}{\left(1+\frac{\kappa r^2}{4}\right)^2}\right] \tag{13c}$$

This reduced form provides a compact and tractable description of blowout boundary dynamics, in which the right-hand side represents the driving sources while the $\Gamma(r)$ coefficients encode geometric and nonlinear corrections. Consequently, the wakefield can be represented as follows:

$$E_z(r,\xi) = \frac{\partial}{\partial \xi}\Phi_{Blowout}(r,\xi) + \frac{\partial}{\partial \xi}\Phi_{RF}(r,\xi) \tag{14}$$

## III. Results and Discussion

In this study, the enhancement of wakefield amplitudes, transverse focusing, and phase-locking in dense magnetized plasmas driven by relativistic electron beams under transverse RF modulation is demonstrated through an integrated theoretical and numerical framework, providing new insights into stabilized blowout regimes for high-gradient plasma-based colliders. To assess and extend the analytical formulation—covering the RF vector potential (Eq. 1), Maxwell's equations and electron motion (Eqs. 2a–c), ponderomotive force (Eq. 3), quasistatic field decompositions (Eqs. 4a–c), plasma electron dynamics (Eq. 5), beam density profile (Eq. 6), cylindrical potentials (Eqs. 7a–c), transverse forces at the blowout radius (Eq. 8), charge continuity (Eq. 9), scalar potential with RF modulation (Eq. 10), the relativistic equation of motion (Eq. 11), and the reduced differential system with modulation coefficients (Eqs. 12–14)—fully electromagnetic particle-in-cell (PIC) simulations were performed with the open-source EPOCH code (version 4.19). These simulations resolve the complete field dynamics, including gyromotion in the external magnetic field, without invoking the quasistatic approximation, thereby capturing multidimensional oscillations and RF-induced phase matching. The computational domain was set in 3D Cartesian geometry with $n_x \times n_y \times n_z = 512 \times 512 \times 2048$ grid cells and a moving window at $\beta_{ph} \approx 0.999c$ in the co-moving coordinate $\xi$, enabling efficient tracking of the highly relativistic electron driver (initial $\gamma \approx 100$). The plasma density was $n_p \approx 10^{18} cm^{-3}$ ($\omega_p = 5.6 \times 10^{13} rad/s$, $k_p = 1.8 \times 10^5 m^{-1}$), with immobile ions providing background space-charge effects. A uniform external magnetic field of B_0≈10 T was applied along z, yielding Larmor radii $r_L \approx 10^{-6}m$ for sheath stabilization. The electron beam had a dual Gaussian profile ($\Theta_r \approx 10\mu m$, $\Theta_z \approx 5\mu m$, peak density $n_b \approx 10^{20} cm^{-3}$), while the RF field was imposed as a transverse wave with normalized amplitudes $A_{0x}^{RF}, A_{0y}^{RF} \approx 0.1$, $\omega_{RF} = 10^{12} rad/s$, $k_{RF} = m^{-1}$, and a phase offset $\varphi = \pi/2$, enabling elliptical polarization. Relativistic corrections, current deposition, and particle–field coupling were handled. Simulations were carried out up to $t_{max} \approx 1\ ps$, producing SDF outputs for fields and particle distributions. Post-processing provided the wakefield amplitude $E_z$, blowout radius $r_b$, and sheath narrowing due to gyration. The results confirmed enhanced wakefield uniformity and amplitude scaling under RF phase-matching in dense magnetized regimes.

The influence of the cyclotron-to-plasma frequency ratio on the radial positioning of electrons within the plasma wakefield is elucidates in Fig. 1, a critical aspect of the model where an external magnetic field induces gyromotion. As this ratio increases, the oscillatory nature of the electron trajectories becomes more pronounced, reflecting enhanced confinement due to the Lorentz force, which stabilizes the blowout regime by counteracting transverse expulsion. The curves suggest a transition from a nearly unmagnetized state to a regime where magnetic effects dominate, leading to a more structured radial distribution that aligns with the theoretical inclusion of $B_{ext}$ in the equation of motion. The amplitude and periodicity of the oscillations are governed by the balance between the Lorentz force induced by the magnetic field and the ponderomotive force arising from the RF driver, with higher ratios amplifying the gyroscopic contribution. The resulting patterns suggest a phase-locked interaction between the electron motion and the wakefield excitation, where the longitudinal coordinate serves as a temporal proxy in the co-moving frame, highlighting the quasistatic approximation's validity in capturing the plasma response. The second sub-figure explores the effect of the RF vector potential amplitude on electron radial dynamics, a key driver in the phase-locked excitation of plasma oscillations. Higher amplitudes of the transverse RF field, as represented by the vector potential, induce greater radial excursions, indicative of stronger ponderomotive forces that modulate electron trajectories. The progressive increase in radial position with amplitude suggests an amplification of the wakefield's transverse focusing, consistent with the model's incorporation of $A_{RF}$ in the ponderomotive force and Maxwell's equations. The observed oscillatory patterns reflect the interplay between the RF field's temporal evolution and the quasistatic plasma response, with the potential for enhanced phase-matching that boosts wake amplitudes in the blowout regime. This analysis highlights the RF field's role in fine-tuning the electron beam dynamics and the resulting radiation polarization properties, offering insights into the optimization of wakefield acceleration in magnetized plasmas.

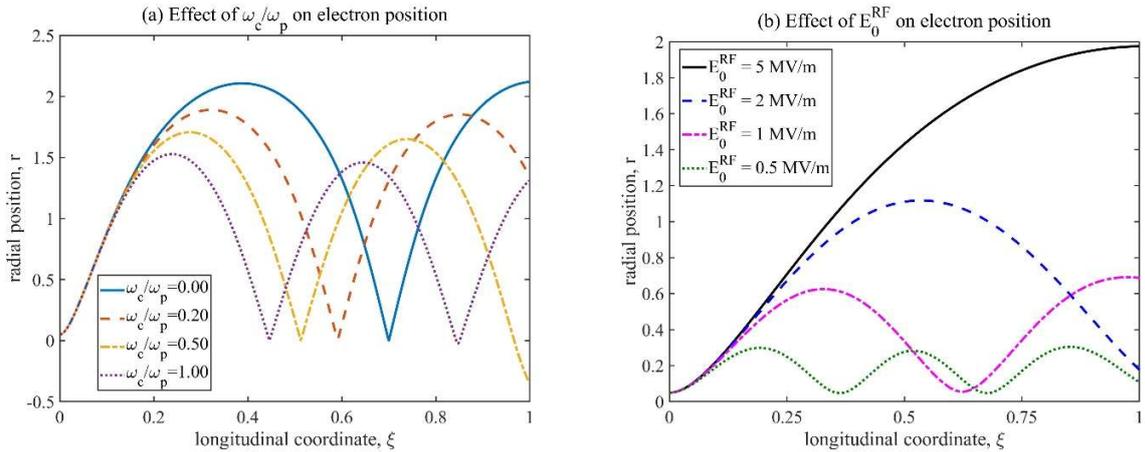

Fig. 1. (a) variation of electron radial position with longitudinal coordinate for different $\omega_c/\omega_p$ ratios, (b) evolution of electron radial position with longitudinal coordinate under varying RF electric field amplitudes.

The modulation of electron radial acceleration by the amplitude of the RF electric field is depicted in Fig. 2, a pivotal driver in the excitation of plasma wakefields within the quasistatic framework. As the field strength increases, the acceleration exhibits amplified oscillatory behavior, reflecting the intensified ponderomotive force that governs transverse electron dynamics, as derived from the RF vector potential in the equation of motion. This enhancement suggests a stronger coupling between the driver and plasma response, potentially leading to heightened wake amplitudes and improved phase-locking, which is critical for efficient energy transfer in the blowout regime. The curves indicate a transition from modest to significant radial excursions, underscoring the RF field's role in shaping the nonlinear plasma oscillations. The subfigure 2b explores the impact of the external magnetic field on radial acceleration, a factor that introduces gyromotion and stabilizes the electron trajectories against blowout. With increasing magnetic field strength, the acceleration profiles display a more confined oscillatory pattern, indicative of the Lorentz force's influence in narrowing the sheath thickness and enhancing transverse focusing. This behavior aligns with the model's incorporation of $B_{ext}$, which mitigates electron expulsion and fosters a more uniform ion channel, thereby supporting higher wakefield amplitudes. The third subfigure of Fig. 2 presents a comprehensive three-dimensional depiction of radial acceleration as a joint function of longitudinal coordinate and magnetic field strength, providing a holistic view of their interplay. The surface plot reveals a complex landscape where acceleration peaks and troughs emerge, driven by the synergistic effects of the RF driver and magnetic confinement. This visualization highlights regions of enhanced acceleration, likely corresponding to optimal phase-matching and gyration-induced stability, which are essential for sustaining the blowout regime. These results on electron radial acceleration, characterized by distinct waves, are consistent with experimental observations, reinforcing the connection between plasma response, wake-field dynamics [27, 28].

The spatial evolution of the parallel current density $J_z$ across the longitudinal and radial dimensions of the plasma wakefield is illustrated in Fig. 3, driven by varying amplitudes of the RF electric field, a central element in the phase-locked excitation mechanism. These visualizations elucidate how increasing RF field strength amplifies the wakefield's current density, enhancing the potential for efficient energy transfer and radiation emission in magnetized plasmas. At the lowest field strength, the current density exhibits a relatively uniform distribution with localized peaks, suggesting a modest plasma response where the ponderomotive force begins to influence electron motion, aligning with the initial stages of wake formation. As the field amplitude increases, the patterns become more pronounced, with sharper peaks and broader oscillatory structures emerging, indicative of enhanced nonlinear interactions between the RF driver and the plasma electrons, which drive the blowout regime's ion channel formation. The progression to higher field strengths reveals the central regions of intense activity expand, and the oscillatory nature intensifies, reflecting the growing dominance of the RF-induced ponderomotive force in modulating electron trajectories. This behavior is consistent with the model's incorporation of $E_0^{RF}$ in the equation of motion, where the transverse electric field contributes to the acceleration and expulsion of electrons, shaping the wakefield's structure. The observed symmetry and periodicity in the current density profiles underscore the quasistatic approximation's validity, where the driver's slow evolution relative to the plasma response allows for a steady-state analysis. The evolution of the parallel current density, in agreement with experimental observations that RF field pulses significantly increase wake-field strength [29, 30].

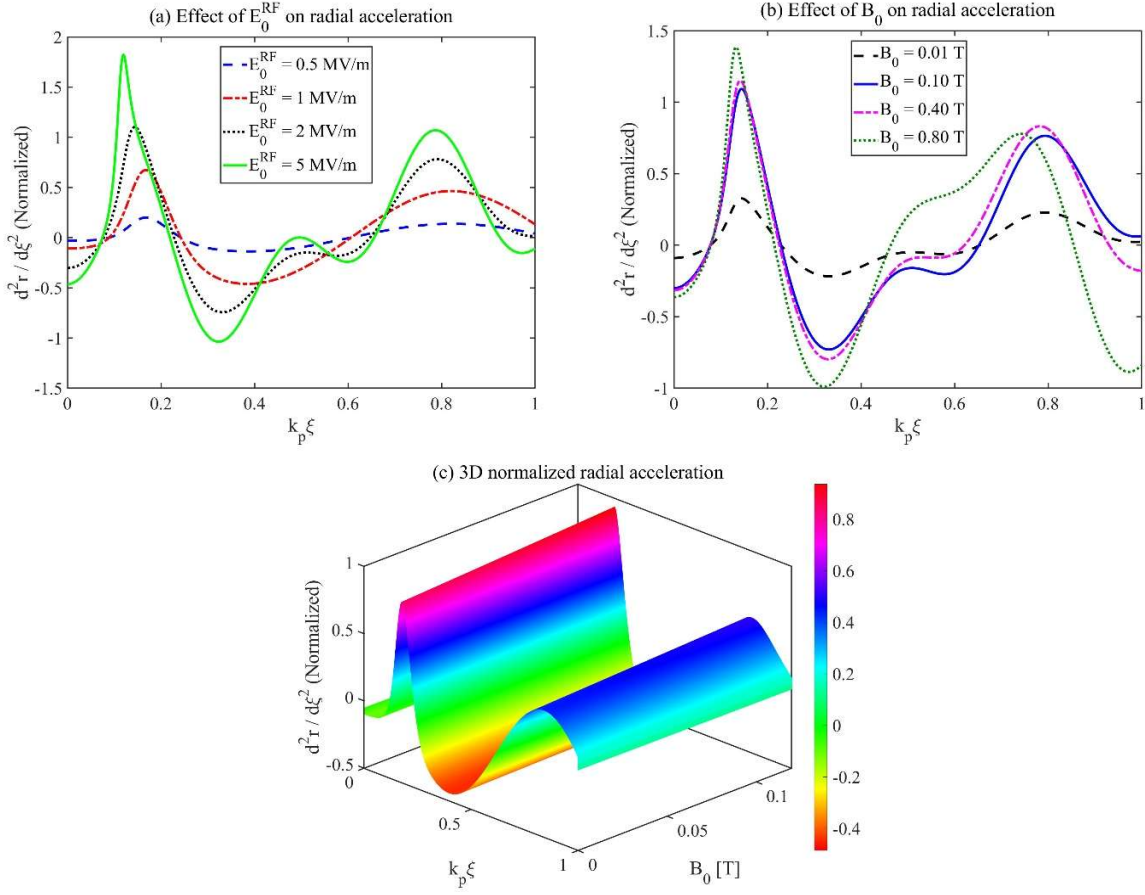

Fig. 2. (a) Normalized radial acceleration of electrons as a function of longitudinal coordinate for varying RF electric field amplitudes, (b) Normalized radial acceleration profiles with longitudinal coordinate under different external magnetic field strengths, (c) Three-dimensional representation of normalized radial acceleration as a function of longitudinal coordinate and external magnetic field.

The variation of the transverse force acting on electrons along the longitudinal coordinate is plotted in Fig. 4, modulated by differing plasma densities, a fundamental parameter influencing wakefield dynamics within the magnetized plasma framework. As the density increases, the force profiles exhibit a progressive shift in amplitude and frequency, suggesting a heightened plasma response where the ponderomotive force, driven by the RF vector potential and external magnetic field, becomes more pronounced. This behavior reflects the model's prediction that higher electron densities enhance the nonlinear coupling between the driver and plasma, leading to a more structured wakefield that supports the blowout regime's ion channel formation. The subsequent three-dimensional representations extend this analysis by mapping the transverse force across both longitudinal and radial dimensions for specific density values, revealing a rich spatial structure that evolves with increasing density. At the lower density, the force distribution shows a relatively smooth variation with localized peaks, indicative of a weaker plasma response where electron expulsion is less pronounced. As the density escalates, the force gradients sharpen, and oscillatory patterns intensify, suggesting a stronger confinement and acceleration mechanism driven by the interplay of the RF-induced ponderomotive force and the magnetic gyration. These results on the transverse force acting on electrons, are consistent with experimental observations, reinforcing the connection between plasma response, wake-field dynamics [31, 32].

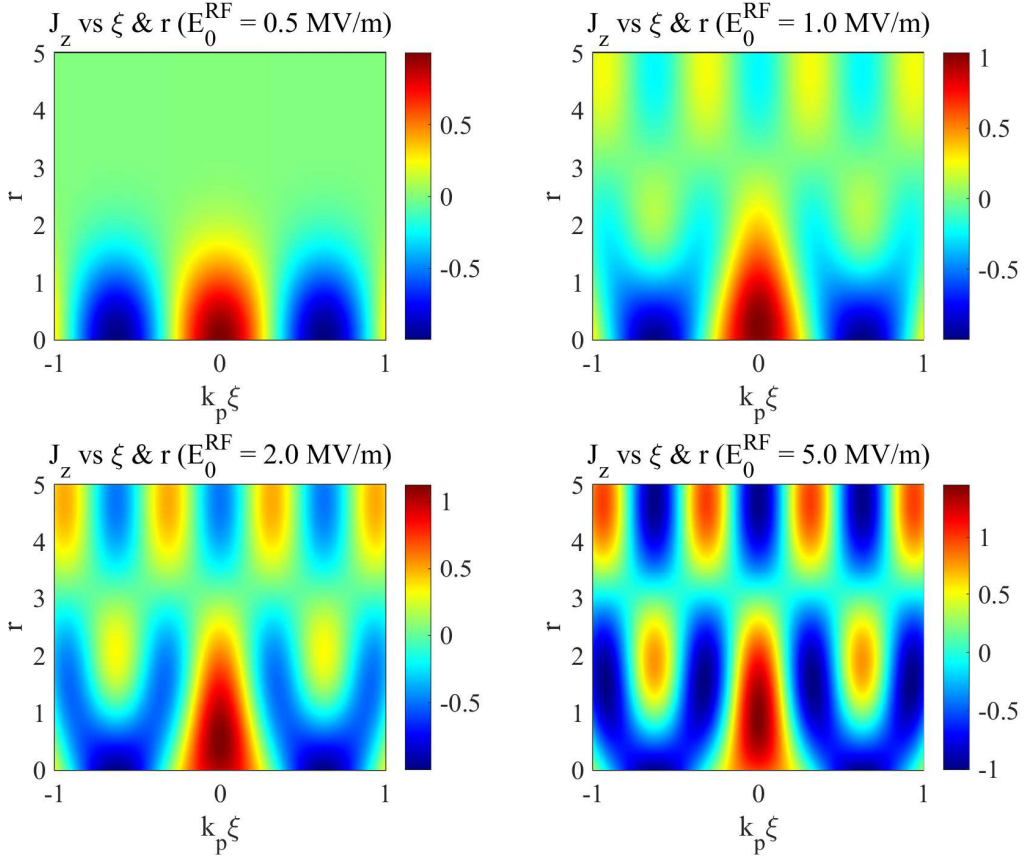

Fig. 3. Spatial distribution of parallel current density $J_z$ with longitudinal and radial coordinates for various RF electric amplitudes.

The variation of the transverse force acting on electrons along the longitudinal coordinate is plotted in Fig. 4, modulated by differing plasma densities, a fundamental parameter influencing wakefield dynamics within the magnetized plasma framework. As the density increases, the force profiles exhibit a progressive shift in amplitude and frequency, suggesting a heightened plasma response where the ponderomotive force, driven by the RF vector potential and external magnetic field, becomes more pronounced. This behavior reflects the model's prediction that higher electron densities enhance the nonlinear coupling between the driver and plasma, leading to a more structured wakefield that supports the blowout regime's ion channel formation. The subsequent three-dimensional representations extend this analysis by mapping the transverse force across both longitudinal and radial dimensions for specific density values, revealing a rich spatial structure that evolves with increasing density. At the lower density, the force distribution shows a relatively smooth variation with localized peaks, indicative of a weaker plasma response where electron expulsion is less pronounced. As the density escalates, the force gradients sharpen, and oscillatory patterns intensify, suggesting a stronger confinement and acceleration mechanism driven by the interplay of the RF-induced ponderomotive force and the magnetic gyration. These results on the transverse force acting on electrons, are consistent with experimental observations, reinforcing the connection between plasma response, wake-field dynamics [31, 32].

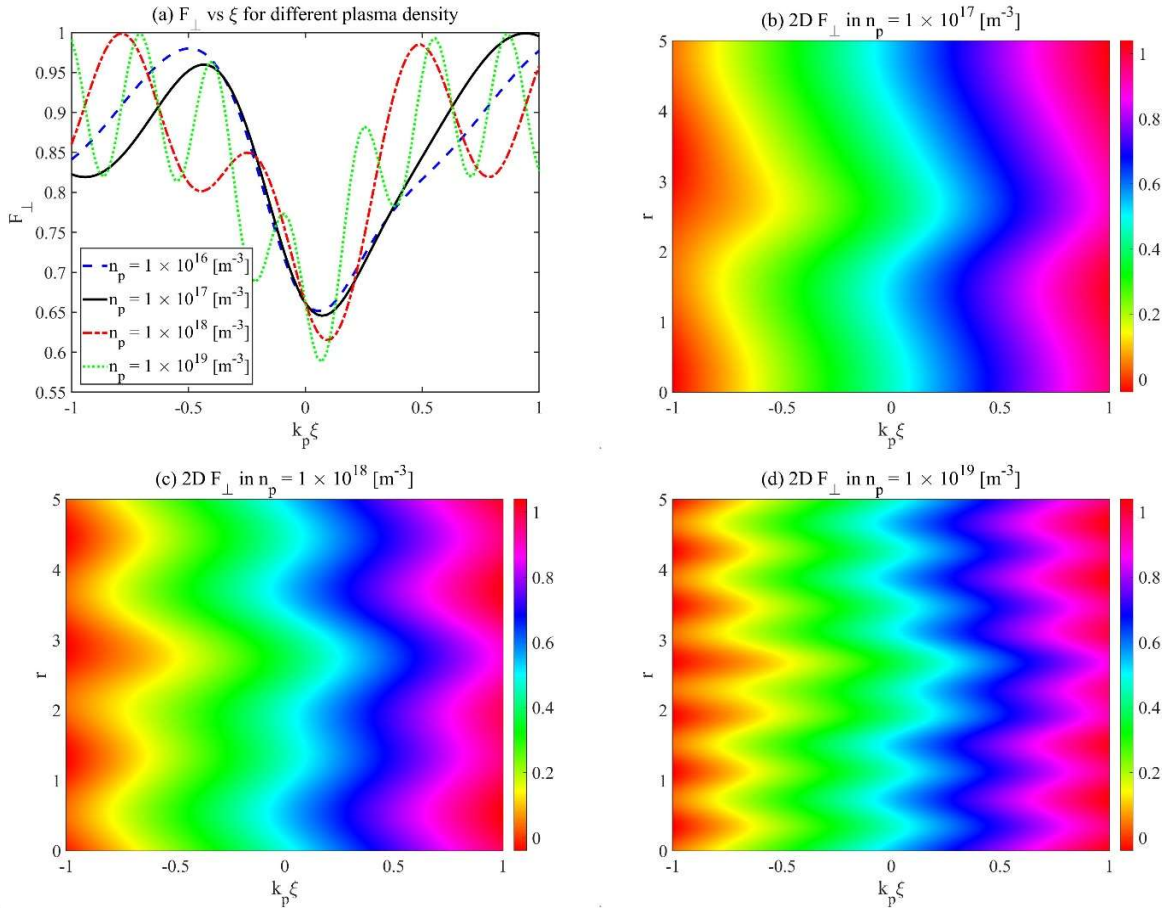

Fig. 4. (a) Normalized transverse force $F_\perp$ as a function of longitudinal coordinate for different plasma densities, (b) Two-dimensional distribution of $F_\perp$ with longitudinal and radial coordinates for various plasma densities.

The longitudinal variation of the transverse force acting on electrons under different RF electric field amplitudes is presented in Fig. 5. As the RF field strength increases, the transverse force grows in magnitude and develops a more pronounced oscillatory structure. This trend directly reflects the strengthening of the ponderomotive force, which originates from gradients in the RF vector potential. At low field amplitudes, the force distribution remains relatively smooth with only modest peaks, corresponding to an early-stage plasma response where the driver weakly perturbs electron trajectories. In this regime, charge displacement is limited, and the wake structure remains shallow. With increasing RF amplitude, however, the force gradients steepen, and oscillatory modulations emerge more clearly, signifying stronger nonlinear coupling between the driver and plasma electrons. The enhanced transverse force not only expels electrons more efficiently but also drives their return motion, processes that are essential for sustaining the blowout regime. This dynamic leads to more effective ion channel formation, where the surrounding positively charged ions provide a self-consistent focusing field that confines electrons and stabilizes the plasma cavity. As a result, higher field strengths support both stronger acceleration fields and improved transverse confinement. The two-dimensional force maps further highlight this progression: while lower RF amplitudes generate smooth distributions indicative of weak perturbations, higher amplitudes produce sharper contrasts and well-defined oscillatory patterns. This evolution demonstrates how the inclusion of $E_0^{RF}$ in the ponderomotive force formulation captures the scaling of electron acceleration and confinement with driver amplitude. Importantly, the consistent normalization applied across the plots emphasizes the relative growth in force features, making it clear how stronger RF fields reshape the transverse force landscape. The longitudinal variation of the transverse force acting on electrons, in agreement with experimental observations that RF field pulses significantly increase wake-field strength [28, 33].

The figure 6 presents a three-dimensional depiction of the scalar potential across the longitudinal and radial dimensions, offering a detailed view of the electrostatic landscape shaped by the plasma wakefield in the presence of the RF driver and external magnetic field. The potential exhibits a pronounced central peak flanked by broader oscillatory features, suggesting a strong localization of charge separation within the ion channel, a hallmark of the blowout regime where electrons are expelled transversely. This distribution reflects the influence of the RF vector potential and the ponderomotive force, which modulate the electron trajectories and contribute to the potential's spatial structure, as derived from the Poisson equation governing $\Phi$. The smooth variation indicates the quasistatic approximation's effectiveness in capturing the steady-state plasma response. The second subfigure extends this analysis by mapping the longitudinal electric field, a critical component of the wakefield responsible for accelerating particles, across the same coordinate space. The field displays a central dip surrounded by oscillatory patterns, indicative of the wake's accelerating and decelerating regions, which are essential for phase-locked energy transfer. This structure aligns with the model's derivation of $E_z$ from the gradient of $\Phi$, enhanced by the RF contribution, and suggests a robust wakefield formation driven by the expelled electron sheath's return under ion space charge. The interplay between the potential and field distributions underscores the nonlinear dynamics of the magnetized plasma, where the RF driver and magnetic confinement collaboratively shape the wakefield.

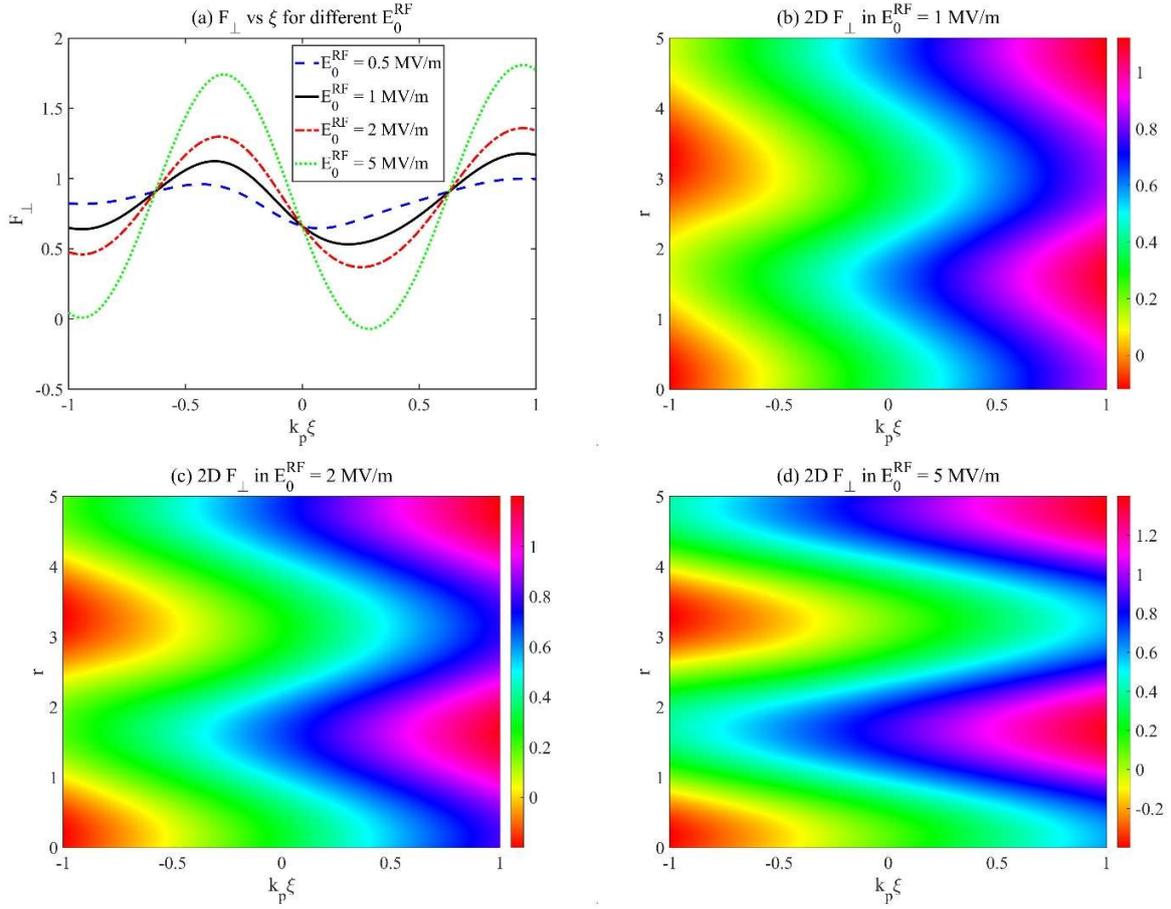

Fig. 5. (a) Normalized transverse force $F_\perp$ as a function of longitudinal coordinate for different RF electric amplitudes, (b) Two-dimensional distribution of $F_\perp$ with longitudinal and radial coordinates for various RF electric amplitudes.

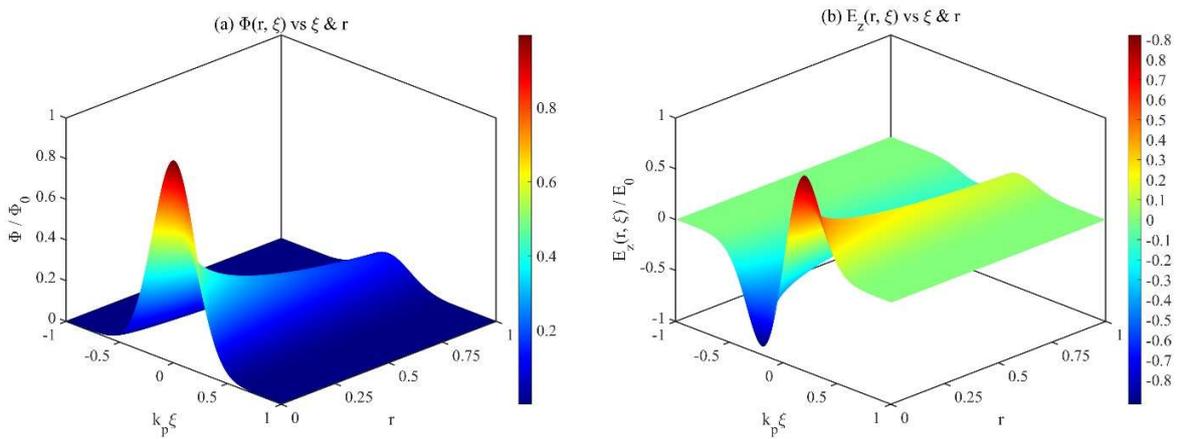

Fig. 6. Spatial variation of normalized scalar potential and electric field with longitudinal and radial coordinates.

The effect of the driver's pulse shape on the longitudinal distribution of the scalar potential is shown in Fig. 7a. The profiles exhibit a central maximum accompanied by oscillatory side structures, with both amplitude and width varying according to the pulse form. A Gaussian pulse produces the sharpest and most symmetric peak, indicating efficient charge separation and a well-defined ion channel, as expected from the smooth pulse gradient of its ponderomotive force. In contrast, Chirped-Gaussian, Ring-Shaped, and Cosh-Gaussian profiles yield broader or more complex structures. The Chirped-Gaussian introduces asymmetry that modifies the phase velocity of the wake, leading to a shifted and less localized potential peak. The Ring-Shaped profile generates a wider ion channel due to its hollow intensity structure, redistributing the ponderomotive force away from the axis and thereby broadening the charge separation. The Cosh-Gaussian produces extended oscillatory tails, reflecting the contribution of its slower-decaying wings, which sustain weaker but longer-range perturbations in the plasma. Together, these differences underscore the sensitivity of wake excitation to the pulse shape of the driver. The corresponding longitudinal electric fields are shown in Fig. 7b. As expected from the derivative of the potential, all profiles display a central dip flanked by oscillatory side peaks, yet the detailed structure depends strongly on the pulse form. The Gaussian case yields a well-balanced acceleration–deceleration pattern, offering efficient and symmetric wakefield formation. In the Chirped-Gaussian case, the temporal frequency variation enhances phase slippage between the driver and the plasma response, resulting in asymmetric field regions and potential improvements in phase-matching under certain conditions. The Ring-Shaped profile generates broader acceleration and deceleration zones, consistent with the extended ion channel and redistributed plasma density around the axis. The Cosh-Gaussian case, with its slower-decaying temporal wings, supports longer wake oscillations that lead to extended acceleration regions but also more diffuse field gradients. These results demonstrate that the temporal structure of the RF driver directly governs the efficiency, symmetry, and phase stability of the wakefield, thereby providing a mechanism for tailoring plasma-based acceleration and radiation generation in magnetized plasmas. The resulting the driver's pulse shape on longitudinal wakefields is consistent with both numerical studies and experimental observations [34].

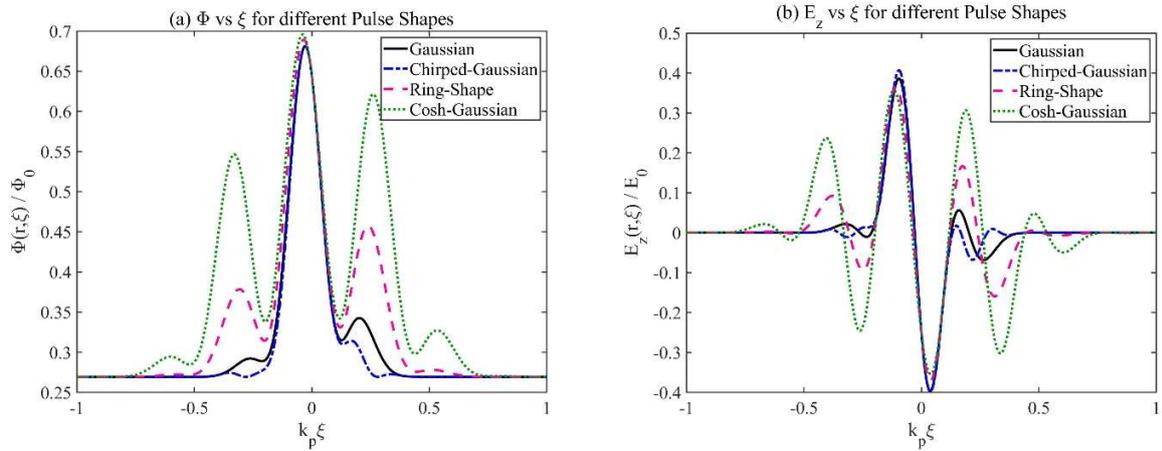

Fig. 7. Variation of normalized scalar potential and electric field with normalized longitudinal coordinate for different pulse shapes.

The figure 8a depicts the longitudinal distribution of the scalar potential under varying polarization angles of the RF vector potential, offering insight into how the elliptical polarization influences the electrostatic structure of the plasma wakefield within the magnetized framework. The potential profiles exhibit a central peak with flanking oscillations, with the amplitude and width subtly modulated by the polarization angle, suggesting a differential impact on charge separation and ion channel formation. The progression from negative to positive phase offsets indicates a shift in the transverse dynamics, where the elliptical nature of the RF field, as governed by the phase φ, alters the ponderomotive force's effectiveness, leading to variations in the wakefield's potential landscape. This behavior aligns with the model's inclusion of φ in Equation 1, highlighting its role in shaping the nonlinear plasma response. The second subfigure of Fig. 8 extends this analysis by mapping the longitudinal electric field along the same coordinate, revealing the acceleration profile influenced by the polarization angle. The field displays a central dip surrounded by oscillatory peaks, with the magnitude and asymmetry evolving as the polarization changes, consistent with the gradient of the scalar potential as derived in the model. The varying phase offsets introduce distinct patterns in the acceleration-deceleration regions, suggesting that the polarization affects the phase-matching between the driver and plasma response, potentially enhancing or suppressing wake amplitudes.

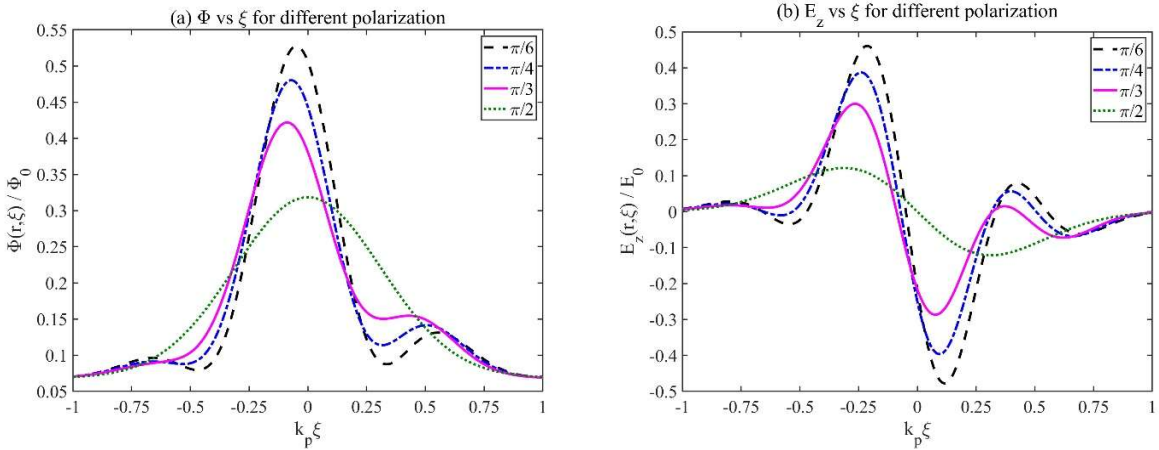

Fig. 8. Impact of polarization states on variation of normalized scalar potential and electric field with normalized longitudinal coordinate.

The influence of the frequency ratio $\omega_{RF}/\omega_p$ on the longitudinal scalar potential is shown in Fig. 9a. The profiles exhibit a central peak surrounded by oscillatory tails, with both amplitude and spatial extent evolving as the frequency ratio increases. At low ratios, the coupling between the RF driver and the plasma response is relatively weak, leading to shallow potential wells and limited ion channel formation. As the ratio grows, a more pronounced central peak emerges, reflecting stronger charge separation and enhanced ion cavity development. This arises from the fact that the RF vector potential directly scales with $\omega_{RF}/\omega_p$, thereby modulating the ponderomotive force that governs electron expulsion in the blowout regime. Larger frequency ratios improve the resonant overlap between the driver and plasma oscillations, increasing the efficiency of energy transfer into the wake. The corresponding longitudinal electric field, depicted in Fig. 9b, displays a central dip flanked by oscillatory peaks whose magnitude and symmetry change systematically with $\omega_{RF}/\omega_p$. Higher frequency ratios intensify these oscillations and broaden the acceleration regions, indicating improved phase-matching conditions. Physically, this reflects the synchronization of RF-driven fields with plasma eigenmodes, which supports stronger and more coherent wake amplitudes, ultimately enhancing acceleration efficiency and wake stability in magnetized plasmas. These results are consistent with experimental observations, reinforcing the connection between plasma response, wake-field dynamics [28, 33]. The effect of RF pulse intensity on the longitudinal scalar potential is illustrated in Fig. 10a. The potential profiles again show a central peak with oscillatory tails, but here the amplitude increases sharply with intensity, indicating stronger charge separation and deeper ion channels. This escalation results from the enhanced ponderomotive force, which scales with the square of the RF field amplitude. As the RF vector potential grows, the nonlinear interaction between the driver and plasma electrons becomes more pronounced, reinforcing the blowout regime and enabling more efficient wake excitation. Figure 10b presents the corresponding longitudinal electric field. The field shows a central dip with oscillatory side peaks whose magnitude and spatial reach expand with intensity, consistent with the gradient of the scalar potential. Higher intensities thus generate broader acceleration zones and deeper decelerating regions. This means that stronger RF pulses increase the peak accelerating field, extend the effective length of the wake, allowing phase-locked particles to gain energy over longer distances.

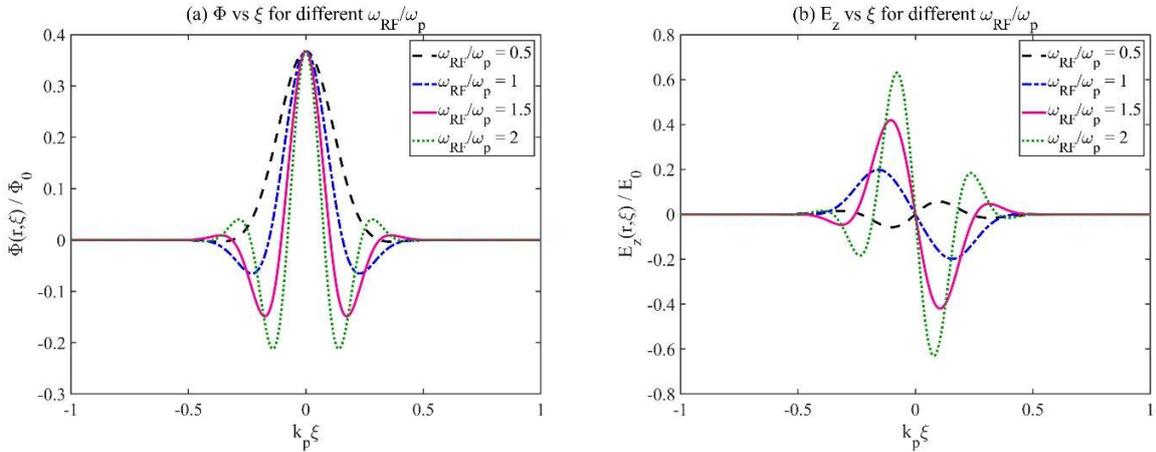

Fig. 9. The role of normalized RF frequency on variation of normalized scalar potential and electric field with normalized longitudinal coordinate.

The longitudinal distribution of the scalar potential under varying driver beam focus sizes is presented in Fig. 11a. The profiles exhibit a central maximum surrounded by oscillatory tails, a typical signature of nonlinear plasma response in the blowout regime. As the focus size increases, both the amplitude and the width of the potential grow. This trend indicates that a broader laser or electron beam spot induces a wider charge separation region, thereby generating an extended ion channel. Physically, this is due to the ponderomotive force, which scales with the transverse intensity profile of the driver: a larger spot size allows the force to act coherently over a wider transverse region, pushing more electrons outward and leaving behind a broader positively charged channel. Consequently, the interaction volume between the driver and plasma electrons is enhanced, leading to stronger and more spatially extended wake structures. Figure 11b shows the corresponding longitudinal electric field. Here, a central dip flanked by oscillatory side lobes is observed, reflecting the gradient of the potential. Both the amplitude and longitudinal span of these oscillations increase with focus size. Larger spots thus give rise to extended accelerating and decelerating phases of the wakefield. This implies that particles can remain in phase with the accelerating field over longer distances, enhancing the efficiency of phase-locked energy transfer, albeit at the cost of a slightly reduced peak gradient due to the energy being distributed across a broader region.

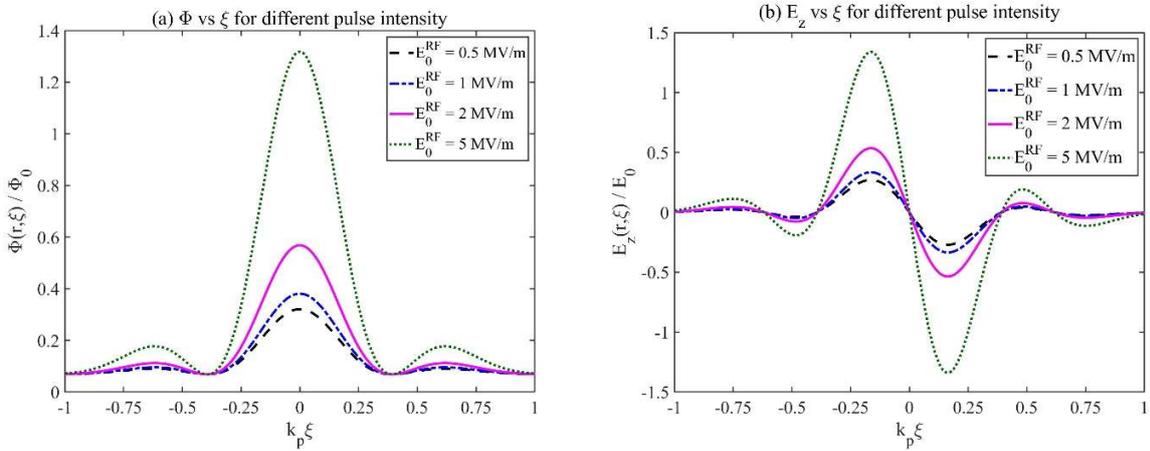

Fig. 10. Variation of normalized scalar potential and electric field with normalized longitudinal coordinate for various RF electric field amplitude.

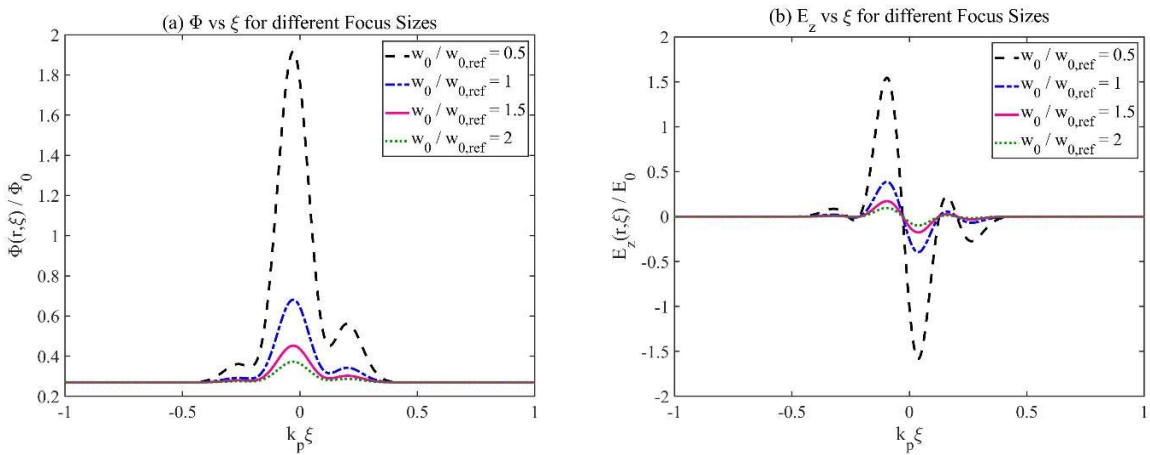

Fig. 11. The impact of focus sizes on variation of normalized scalar potential and electric field with normalized longitudinal coordinate.

The effect of the external magnetic field is displayed in Fig. 12. The scalar potential profiles show a central peak with oscillations, but unlike the focus-size case, the profiles here become more confined and sharper with increasing magnetic field strength. The higher confinement results from the Lorentz force, which induces stronger electron gyration, thereby counteracting the transverse expulsion caused by the ponderomotive force. As a result, electrons are guided more efficiently along helical trajectories, tightening the blowout cavity and reinforcing charge separation near the axis. Figure 12b illustrates the corresponding longitudinal electric field. As the magnetic field strengthens, the dip at the center becomes deeper and the oscillatory peaks more pronounced, reflecting the sharper gradients in the scalar potential. This produces more localized acceleration regions with higher peak fields, accompanied by narrower decelerating zones. Physically, the external magnetic field introduces an additional restoring mechanism that enhances wakefield stability, leading to a more focused accelerating structure capable of sustaining stronger, phase-locked acceleration of injected particles.

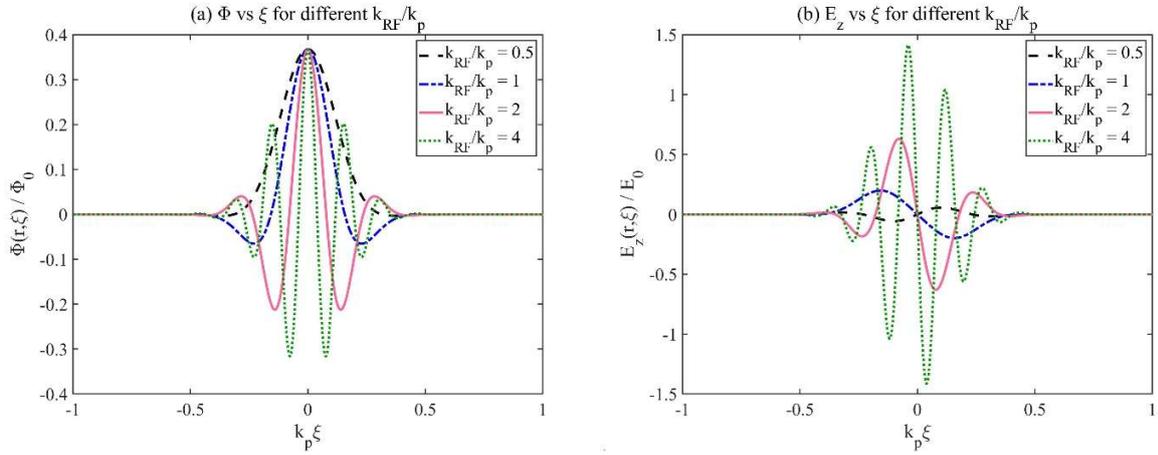

Fig. 12. Variation of normalized scalar potential and electric field with normalized longitudinal coordinate for different normalized RF wave numbers.

Figure 13 demonstrates the dependence of the scalar potential on the beam-to-plasma density ratio. As this ratio increases, the central peak of the potential deepens, and the surrounding oscillatory tails become more pronounced. This occurs because a denser driver beam carries more space charge, intensifying both the ponderomotive expulsion of electrons and the nonlinear plasma response. The result is a stronger blowout cavity with enhanced charge separation and a more robust ion channel. In Fig. 13b, the corresponding longitudinal electric field reveals deeper central dips and larger-amplitude oscillatory peaks as the density ratio rises. The acceleration and deceleration regions both become broader, indicating that the wakefield extends over a larger spatial volume. Physically, a higher beam density injects more energy into the plasma, amplifying wake excitation and supporting phase-locked energy transfer over a wider domain. However, this also increases nonlinearity, which can lead to stronger wave-breaking effects if not properly controlled. Thus, the growth of both scalar potential and longitudinal field with increasing beam density highlights the dual role of intense drivers: they enhance acceleration capability while simultaneously demanding stronger stabilization mechanisms to maintain wake coherence. These results are consistent with theoretical predictions that aligns with experimental findings [35, 36].

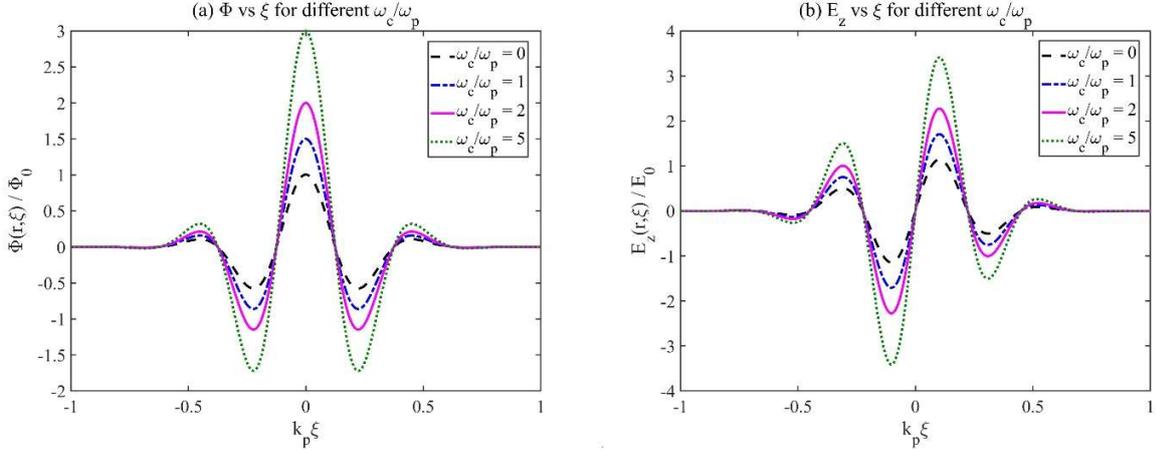

Fig. 13. The effect of cyclotron frequency on variation of normalized scalar potential and electric field with normalized longitudinal coordinate.

In the upper row of figure 14, the panels present the longitudinal electric field distribution as the plasma density is varied. At the lowest density, the field exhibits a relatively smooth longitudinal dependence with moderate amplitude. The axial variation is gentle, and the radial decay is gradual, indicating that the restoring force of the plasma is weak, and electron displacement remains limited. When the density is increased, the field amplitude rises, and the longitudinal profile develops sharper gradients around the central region. The radial extent of the field also becomes more confined, reflecting the fact that higher plasma density enhances the plasma frequency and strengthens the electrostatic restoring force, which in turn compresses the wake structure toward the axis. The oscillatory structure in the longitudinal direction is more pronounced, and the radial confinement is clear, illustrating how dense plasmas support intense but spatially compressed wakefields. The progression across the density values therefore reveals a continuous transition from weak, extended fields at low density to strong, narrow, and sharply defined fields at high density. The lower row shows the influence of the parameter $\kappa$ on the longitudinal electric field. For the smallest $\kappa$, the distribution is dominated by a smooth central region with only weak oscillatory signatures. The longitudinal variation is mild, and the field remains largely uniform across the radial direction. As $\kappa$ increases, oscillations along the longitudinal axis become more evident, producing alternating positive and negative regions of the electric field. These oscillations extend radially outward, and their periodicity becomes more regular and pronounced. At intermediate $\kappa$, several distinct oscillatory bands appear, each separated by narrow regions of opposite sign, indicating stronger modulation of the plasma response. At the largest $\kappa$, the longitudinal field is highly oscillatory, with densely packed alternating bands spanning the axis to the outer radial edge. The field variations are sharp, and the contrast between maxima and minima is strong, showing that large $\kappa$ values drive the system into a regime dominated by rapid spatial modulations.

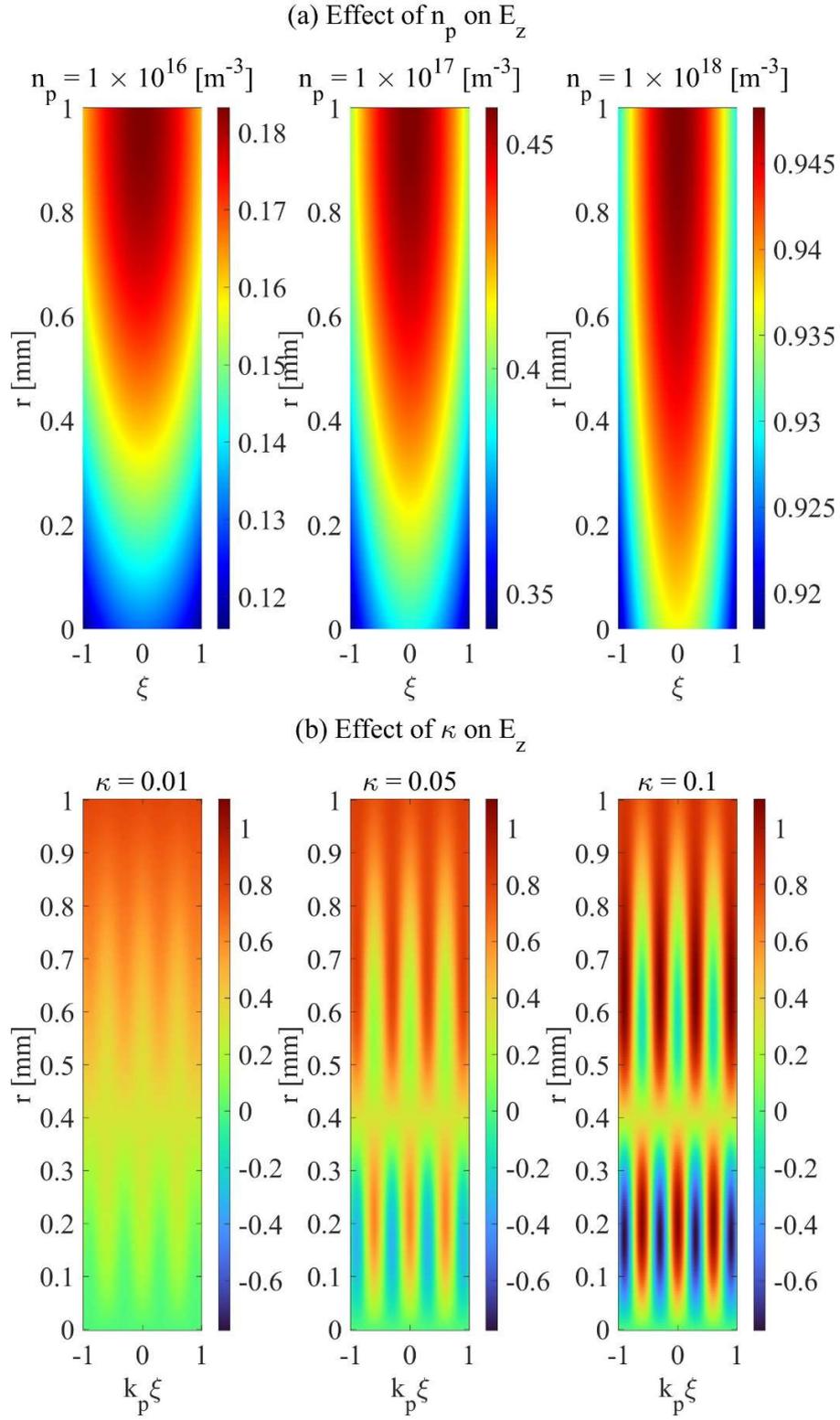

Fig. 14. Effect of (a) plasma density, and (b) the strength modulate parameter on the radial and longitudinal evaluation of $E_z$.

## IV. Conclusions

The present study investigates the modulation of plasma wakefields in dense magnetized plasmas driven by relativistic electron beams under the influence of externally applied transverse RF fields. A self-consistent theoretical framework is developed, incorporating the RF vector potential into Maxwell's equations coupled with the relativistic equations of motion for plasma electrons, and is extended and validated through fully electromagnetic three-dimensional particle-in-cell simulations. This combined analytical–numerical approach enables a detailed examination of how wakefields are systematically amplified, reshaped, and stabilized by the interplay between RF excitation and magnetic confinement. The presented results establish a consistent picture of how relativistic beam-driven wakefields evolve under the combined influence of transverse RF modulation and external magnetic confinement in dense plasma environments. The simulations confirm that plasma density primarily determines field amplitude and radial confinement, producing a transition from broad, weakly modulated wakes to strongly localized, sharply defined structures at higher densities. The modulation parameter $\kappa$ introduces progressive oscillatory complexity, with low values yielding smooth distributions and high values generating dense multiband longitudinal patterns. RF field amplitude directly scales radial excursions, transverse force gradients, and current density modulation, enhancing sheath narrowing and ion channel sharpening. Variations in polarization, pulse shape, and frequency ratio produce distinct asymmetries, broadenings, or phase shifts in the scalar potential and longitudinal fields, reflecting the sensitivity of wake formation to driver configuration. External magnetic fields enforce gyromotion that counteracts ponderomotive expulsion, leading to sheath stabilization and enhanced focusing. Across all parameter regimes, the interplay of Lorentz and ponderomotive forces consistently reshapes the wake into a confined, oscillatory structure with extended acceleration phases. These findings establish a parameter-dependent picture of wakefield evolution in magnetized, RF-driven beam–plasma systems. The study identifies the physical mechanisms responsible for amplitude growth, oscillatory complexity, and sheath stabilization, providing predictive guidelines for optimizing plasma-based accelerators.

**Acknowledgment**

This work did not receive any specific grant from funding agencies in the public, commercial, or not-for-profit sectors.

**Author Contributions**

Ali Asghar Molavi Choobini was responsible for conceptualization, data curation, formal analysis, investigation, methodology, writing the original draft, and review and editing, all with equal contributions.

M. Shahmansouri contributed equally to investigation, project administration, supervision, validation, and review and editing of the manuscript.

## Data availability statement

The data that support the findings of this study are available from the corresponding author upon reasonable request.

## Competing interests

The authors declare no competing interests.